\documentclass[twocolumn,preprintnumbers,amsmath,amssymb,aps,prl]{revtex4}
\usepackage{graphicx}

\usepackage{tikz}

\newcommand{\tikzcircle}[2][black,fill=black]{\tikz[baseline=-0.5ex]\draw[#1,radius=#2] (0,0) circle ;}%
\newcommand{\tikzcirclew}[2][black,fill=white]{\tikz[baseline=-0.5ex]\draw[#1,radius=#2] (0,0) circle ;}%

\begin{document}

\title{The Inner Phases of Colloidal Hexagonal Ice}
 
\author{A. Lib\'{a}l$^{1,2}$, C. Nisoli$^{1}$, C.J.O. Reichhardt$^{1}$, and C. Reichhardt$^{1}$}
\affiliation{$^{1}$Theoretical Division and Center for Nonlinear Studies,
Los Alamos National Laboratory, Los Alamos, New Mexico 87545, USA}
\affiliation{$^{2}$Mathematics and Computer Science Department, Babe{\c s}-Bolyai University, Cluj, Romania 400084}

\date{\today}

\begin{abstract}
Using numerical simulations that mimic recent experiments on hexagonal colloidal ice, we show that colloidal hexagonal artificial spin ice exhibits an inner phase within its ice state that has not been observed previously.  Under increasing colloid-colloid repulsion, the initially paramagnetic system crosses into a disordered ice-regime, then forms a topologically charge ordered state with disordered colloids, and finally reaches a three-fold degenerate, ordered ferromagnetic state.  This is reminiscent of, yet distinct from, the inner phases of the magnetic kagome spin ice analog. The difference in the inner phases of the two systems is explained by their difference in energetics and frustration. 
\end{abstract}

\maketitle

\vskip 2pc

{\it Introduction.} Artificial spin ice (ASI) systems have been attracting increasing interest
as frameworks for studying
frustration or degeneracy and for revealing emergent exotic behaviors \cite{1,2,3}. 
Among the most studied ASIs are
nano-scale magnets arranged in square  \cite{1,4,5,6,7,8,9}, hexagonal
\cite{1,10,11,12,13}, or other geometries \cite{14,15,16,17,18,19,20,N}. 
Each individual magnet behaves like a binary spin degree of freedom
and adopts, to a nearest-neighbor (NN) approximation,
an ice-rule configuration that minimizes the topological charge $q_n$, defined as
the absolute difference between the $n$ spins pointing in and
$n_o$ spins pointing out of each vertex.
This configuration can be ordered, as in square ice \cite{1,4}, or it
can be a disordered manifold with non-zero entropy density.
Moving beyond the NN approximation, inner phases and transitions appear
within the disordered ice manifold.
In hexagonal ice,
each vertex is surrounded by $v_n=3$ spins and the ice rule corresponds to
$n=1$, $n_o=2$ ($q_n=2n-v_n=-1$) or $n=2$, $n_o=1$ ($q_n=+1$).
In disordered hexagonal or kagome ice, inner phases
corresponding to charge ordering (CO) within spin disorder (the ``spin ice II'' or SI2 phase)
and to long range order (LRO)
have been reported
\cite{21,22} and experimentally investigated \cite{10,12,23,24}.

\begin{figure}
\includegraphics[width=3.3in]{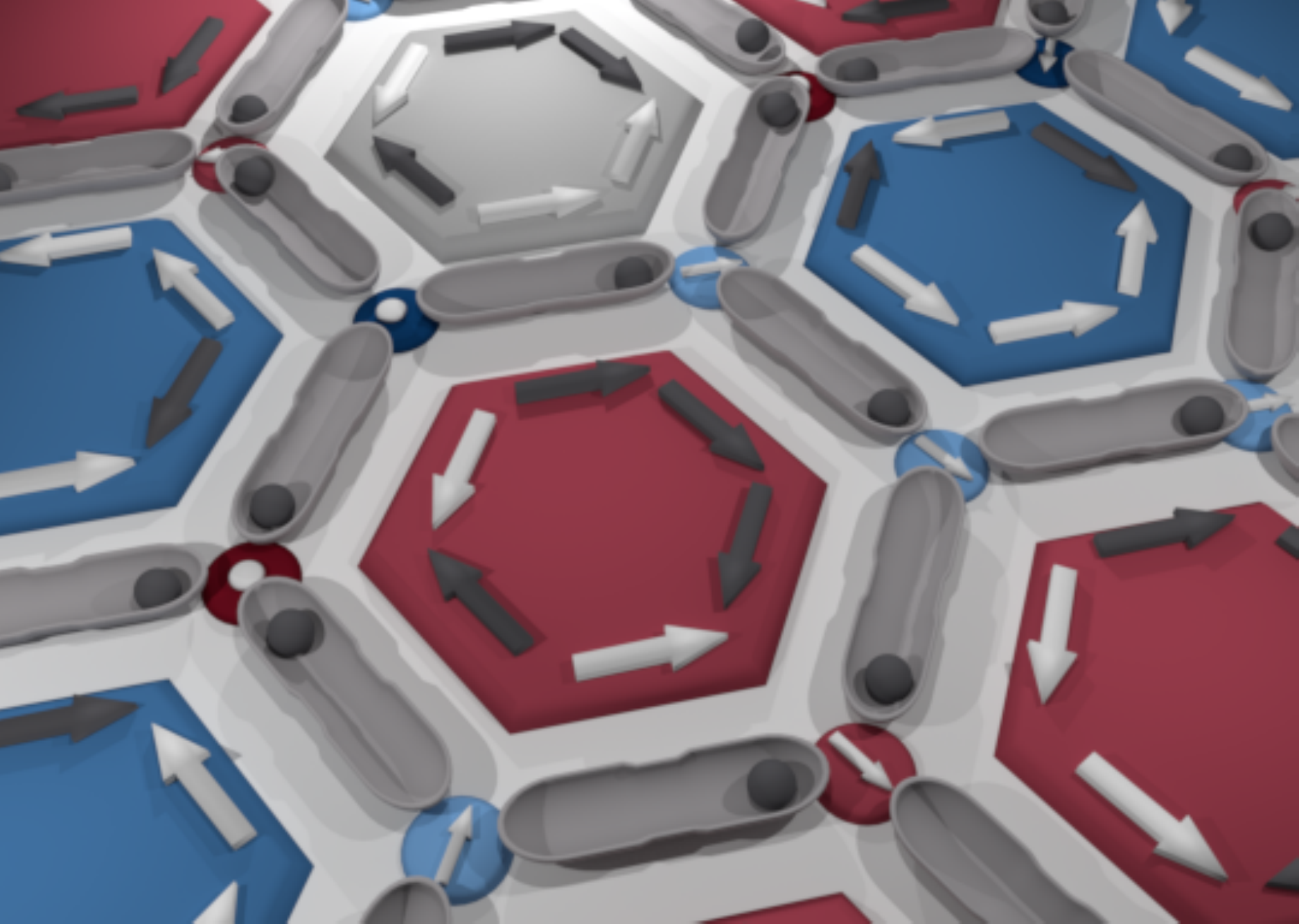}
\caption{Schematic of the particle-based
  hexagonal artificial spin ice.
  Each double well trap (light grey) holds a single 
  paramagnetic colloid (dark grey dots).
  The hexagonal plaquettes contain arrows indicating
  the pseudospin $\vec \sigma_i$ of the adjacent traps, colored according to
  the chirality
  $\chi_i=+1$ (clockwise, dark grey) or $\chi_i=-1$ (counter-clockwise, white).
  The plaquettes are colored according to their net spin chirality $\chi$:
  clockwise (red), counter-clockwise (blue), or achiral (grey).
  Colored disks are guides to the eye and indicate the vertex type:
  $n=0$ or 0-in (dark blue), $n=1$ (light blue), $n=2$ (light red), and
  $n=3$ (dark red); arrows (of length 2) or dots (of length 0) on the
  disks indicate the vectorial sum $\vec s_i$ of the pseudospins adjacent to each vertex.
}
\label{fig:1}
\end{figure}

Another interesting class of ASIs that resemble water ice
consists of an array of double-well traps  
that each capture one particle, as illustrated in Fig.~\ref{fig:1}.
The traps are arranged in a square or hexagonal ice geometry with
$v_n=4$ or $v_n=3$ traps, respectively,
around each vertex, and the particle-particle interactions are {\it repulsive}.
Particle-based ASIs have been studied numerically for colloids 
\cite{26,28,29,30,31,32,33}, skyrmions \cite{N2},
and vortices in type-II superconductors \cite{27},
and have been realized experimentally
in superconductors \cite{37,38,39,40} and   
for paramagnetic colloids on grooved surfaces \cite{34,35,36}.

Particle-based ASIs can be described in the same way as
magnetic spin ices by defining
a pseudospin $\vec \sigma_i$ lying along the trap axis and pointing toward the
particle \cite{26,27,28}.
Although the particle and magnetic ASIs
differ greatly both in energetics and frustration,
both obey the ice rules at low energy, though for different reasons \cite{28}.
Consider for definiteness $v_n=3$ magnetic and colloidal
hexagonal or kagome ASIs, which both obey the ice rule
through a local minimization of $q_n$.
The NN energy $E_n$ of magnetic dipoles
impinging into a vertex is  proportional to the square of the charge,
$E_n\propto q_n^2$. Thus the ice rule ($q_n=\pm1$ allowed and
$q_n=\pm3$ forbidden)
is enforced locally by energy minimization.
For the colloids, $E_n\propto n(n-1)$,
and the vertex energetics favors large negative charges ($q_n=-3$).
Thus the ice rule obeying particle-based ASI
minimizes the {\it global} energy of the system
rather than the local energy, as in its magnetic counterpart \cite{28}.   

An open question is how far the similarities between the particle-based and magnetic
ASIs extend.  Here we explore the types of
ice rule states that occur in particle-based ASI
and how they relate to the CO and LRO of magnetic systems.
We simulate paramagnetic colloids held gravitationally in
double-well etched grooves, similar to those used in
recent experiments \cite{35,36}.
The colloids repel each other with interaction $\propto B^2/r^4$
when a perpendicular magnetizing field $\vec B$ is applied.
The strength of the spin-spin interactions can be tuned
easily by varying $\vec B$, permitting us to
map different ice phases. 
For weak interactions the system is in a paramagnetic state
containing $q_n=\pm 3$ charges,  
but we find that for higher interaction strengths the system enters an ice-rule phase
with no $q_n=\pm 3$ charges.
As $\vec B$ is increased further, 
a charge ordering regime emerges in which
the colloids remain disordered,
and the order gradually increases until the system forms
domains of three-fold symmetric ferromagnetic order. 

{\it Simulation--}
We conduct Brownian dynamics simulations of the
particle-based hexagonal ASI
comprised of $N_p=2700$ magnetically interacting colloids
with diameter $1\mu$m placed
in an array of $N_t=2700$ etched double-well grooves. 
There are $N_{pl} = 900$ hexagonal
plaquettes of side $a_h=3\mu$m arranged on a $15\times20$ lattice with dimensions 
of $135\mu$m $\times 155.88\mu$m, and we use periodic boundary
conditions in both the $x$ and $y$ directions.
Each plaquette is surrounded by six double-well traps of length
$a_t=2.8\mu$m, as illustrated in Fig.~\ref{fig:1},
giving  a total of $N_v=1800$ vertices. 
A confining spring force ${\bf F}_{c1}$
acts perpendicularly to the elongated direction of the trap,
and each end of the trap contains a confining parabolic attractive well
exerting force ${\bf F}_{c2}$
with a spring constant of $2.2$pN/$\mu$m
representing the gravitationally induced attraction.
A repulsive harmonic barrier ${\bf F}_h$
of magnitude
$2.11$ pN corresponding to a barrier of height 3.32$\mu$m
separates the attractive wells.
The combined substrate forces are written as
${\bf F}_s={\bf F}_{c1}+{\bf F}_{c2}+{\bf F}_h$.
Magnetization of the colloids in the $z$ direction produces
a repulsive particle-particle
interaction force ${\bf F}_{pp}(r)=A_c{\bf \hat r}/r^4$
with $A_c=3\times 10^6\chi_m^2 V^2 B^2/(\pi\mu_m)$
for colloids a distance $r$ apart.
Here $\chi_m$ is the magnetic susceptibility,
$\mu_m$ is the magnetic permeability,
$V$ is the colloid volume,
$B$ is the magnetic field in mT,
and all distances are measured in $\mu$m.
For the paramagnetic colloids in Ref.~\cite{35,36},
this gives
$|{\bf F}_{pp}|=6.056$ pN
for $r=3\mu$m at 
$B=40$ mT, the maximum field we consider.
The dynamics of colloid $i$ are obtained using
the following discretized overdamped equation of motion:
\begin{equation}
  \frac{1}{\mu}\frac{\Delta {\bf r}_i}{\Delta t} = \sqrt{\frac{2}{D\Delta t}}k_B T N[0,1] + {\bf F}_{pp}^i + {\bf F}_s^i
\end{equation}
where the  diffusion constant $D=36000\mu$m$^2$/s,
the mobility $\mu = 8.895 \mu$m pN/s, the
simulation time step $\Delta t = 1$ms, and where N[0,1] is a Gaussian distributed
random number with mean 0 and variance 1.
The first term on the right is a thermal force consisting of Langevin kicks of magnitude
$F_T=0.95$ pN corresponding to a temperature of $t=20^\circ$C. 
Each trap is initially filled with a single colloid placed in a randomly chosen well.
We increase $B$ linearly from $B = 0$ mT
to $B = 40$ mT, consistent with the experimental range \cite{34}.
Unless otherwise noted, we average the results
over $100$ simulations performed with different random seeds.
Around a hexagonal plaquette, there is a pseudospin to the right and left of
each vertex.  Considering the set of right pseudospins,
we define the pseudospin chirality $\chi_i=+1$ if the pseudospin is pointing
toward the vertex and -1 otherwise.
The net chirality of each
plaquette is $\chi=\sum_{i=1}^6\chi_i/6$, as illustrated in Fig.~\ref{fig:1}.
We assign a chirality direction (clockwise or counter-clockwise) to each plaquette based 
on the sign of $\chi$. In the case of achiral $\chi=0$
plaquettes, when possible we assign an 
effective biasing field $\vec F_{b}$ to each plaquette representing the
in-plane biasing field that would have produced the same spin ordering.

\begin{figure}
\includegraphics[width=3.5in]{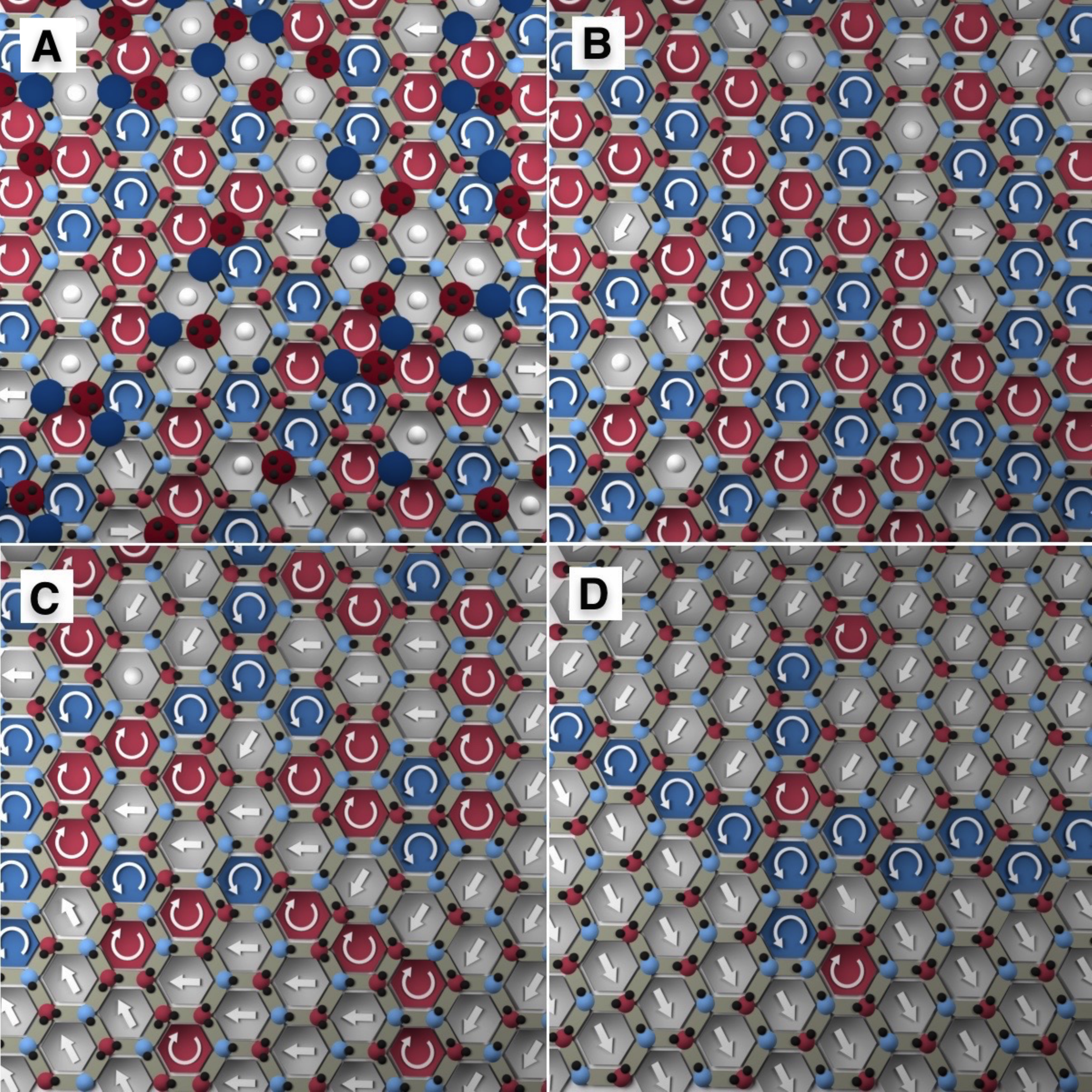}
\caption{Images of a small portion of the sample colored as in Fig.~\ref{fig:1}, where 
  the pseudospin arrows are replaced by an arrow
  indicating the plaquette chirality direction or
  effective biasing field $F_{b}$ for chiral and achiral plaquettes, respectively.
  Dots indicate that no $F_b$ value can be assigned.
  (a) Paramagnetic (PM) phase at $B=0$ mT.
  Large red and blue disks indicate $q_n=\pm 3$ vertices with
  $n=3$ and $n=0$, respectively.
  (b) Charge-free (ICE) phase at $B=13.2$ mT containing only $q_n=\pm 1$ vertices.
  (c) Partially charge ordered (PCO) phase at $B=24$ mT with domains of
  charge and spin ordered vertices and plaquettes.
  (d) Ferromagnetic (FM) phase at $B=40$ mT containing a grain boundary.
  The system contains a second grain boundary with complementary chirality (not shown).}
\label{fig:2}
\end{figure}

{\it Results--}
In Fig.~\ref{fig:2}, we illustrate
the four phases exhibited by the system.
In the paramagnetic (PM) phase, shown at $B=0$ mT in Fig.~\ref{fig:2}(a),
$q_n=\pm 3$ charges are present.  At $B =  13.2$ mT in
Fig.~\ref{fig:2}(b), we find a
charge-free (ICE) phase containing no $q_n=\pm 3$ charges. 
Here all the vertices obey the ice rules but there is no
CO or ferromagnetic ordering.
In Fig.~\ref{fig:2}(c) at $B = 24$ mT,
a partially charge ordered (PCO) phase appears in which
the vertices obey the ice rules and
some CO arises in the form of
$n=2$ vertices surrounding $n=1$ vertices and vice-versa.
At $B=40$ mT
in Fig.~\ref{fig:2}(d),
there is pronounced CO
and chiral plaquettes only exist along grain boundaries.
This ferromagnetic (FM) phase contains two domains
with net effective biasing field $\vec F_b \neq 0$.
Since there are six possible $\vec F_b$ orientations,
the FM phase often exhibits domains and
grain boundaries.       

\begin{figure}
  \includegraphics[width=\columnwidth]{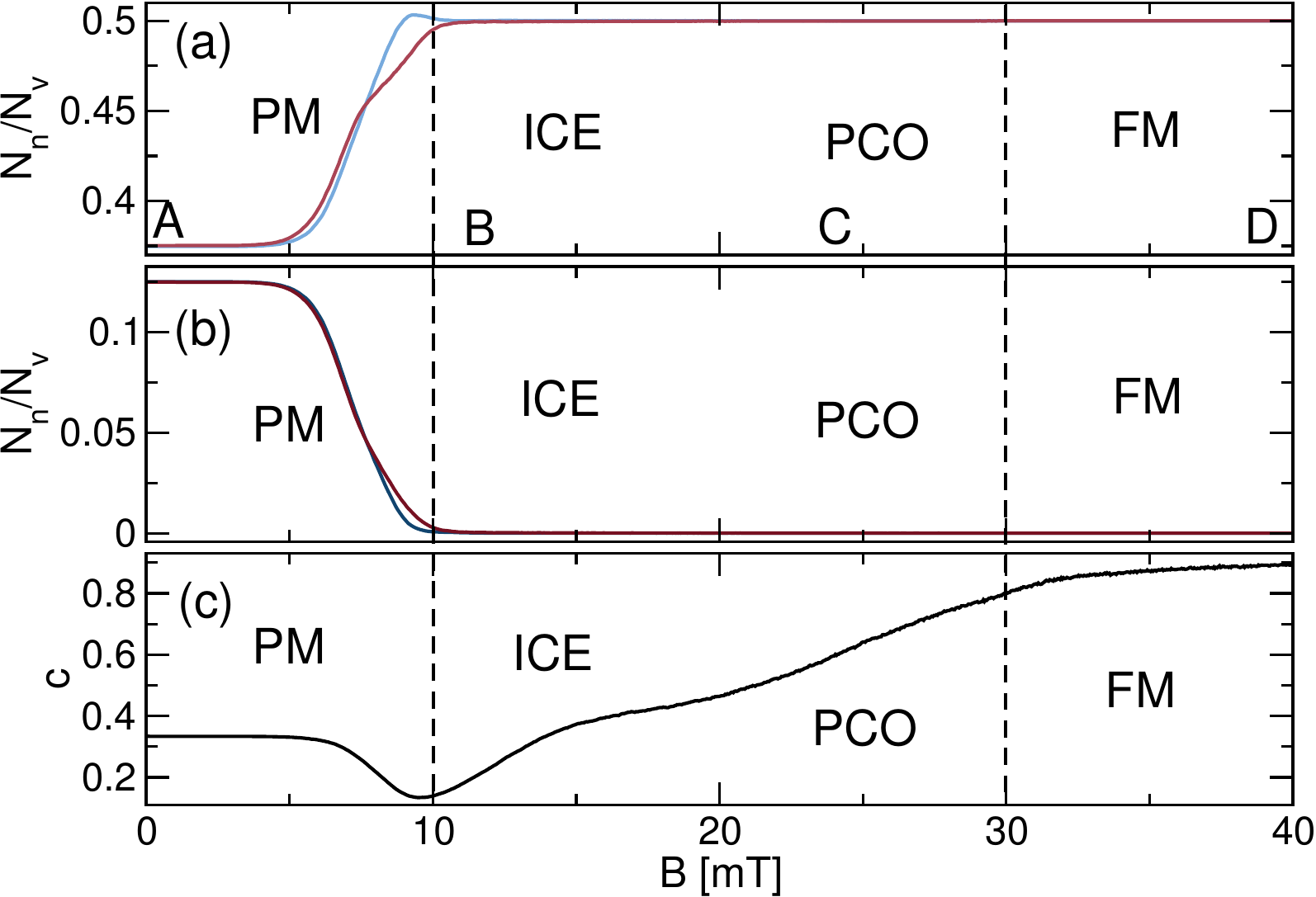}
\caption{(a) Fraction of ice rule obeying $q_n=\pm 1$ vertices $N_n/N_v$
  vs $B$,  where $N_n$ is the number of vertices with $n$ ``in''
  pseudospins.
  Blue: $n=1$, $q_n=+1$; red: $n=2$, $q_n=-1$.
  PM: paramagnetic; ICE: charge free;
  PCO:  partially charge ordered;  FM: ferromagnetic.
  Labels A to D indicate the $B$ values
  illustrated in Fig~\ref{fig:2}.
  Dotted lines indicate approximate transition locations; the
  ICE-PCO crossover field is not well defined.
  (b) Fraction of $q_n=\pm 3$ charged vertices vs $B$.
  Blue: $n=0$ with $q_n=-3$; red: $n=3$ with $q_n=+3$.
  The charged vertices disappear above
  $B = 10$ mT at
  the PM-ICE transition.
  (c) Charge ordering parameter $c$
  vs $B$.
}

\label{fig:3}
\end{figure}

In Fig.~\ref{fig:3}(a) we plot the fraction $N_n/N_v$ of ice rule obeying vertices
with $q_n=\pm 1$ versus $B$, and in 
Fig.~\ref{fig:3}(b) we show the corresponding fraction of
$q_n=\pm 3$ vertices versus $B$. 
For $B=0$ mT when the colloids do not interact with each other, the
vertices are randomly distributed, giving
$N_0/N_v = N_3/N_v = 1/8$ and $N_1/N_v = N_2/N_v = 3/8$.
As $B$ increases,
there is a transition to $N_0/N_v=N_3/N_v=0$ near
$B=10$ mT when the system enters an ice rule obeying state.

We introduce a charge order parameter 
$c =- \frac{1}{N_v}\sum_{i=0}^{N_v}\Big( \frac{1}{q_n^i}\sum_{ i \in \partial_i} q_n^j\Big)$
to measure the charge-charge correlation among NN vertices.
In a random system, $c = 1/3$, while $c=1$ in a CO state.
In Fig.~\ref{fig:3}(c) we plot $c$ versus $B$  showing that
in the PM phase, $c=1/3$,
and at the PM-ICE transition, $c$ drops.
In the ICE phase ($10$ mT $< B < 15$ mT),
$c<0.4$ and the ice rule is obeyed.  For $15$ mT $< B < 30$ mT in the PCO phase,
$c$ gradually increases, saturating
to $c=0.9$ in the FM phase for $B > 30$ mT.
Here $c<1.0$ since the FM grain boundaries disrupt
the CO, as shown in Fig.~\ref{fig:2}(d) for $B = 40$ mT.          
In magnetic ASI, a
SI2 phase appears when the system 
has CO but no spin ordering.

\begin{figure}
  \includegraphics[width=\columnwidth]{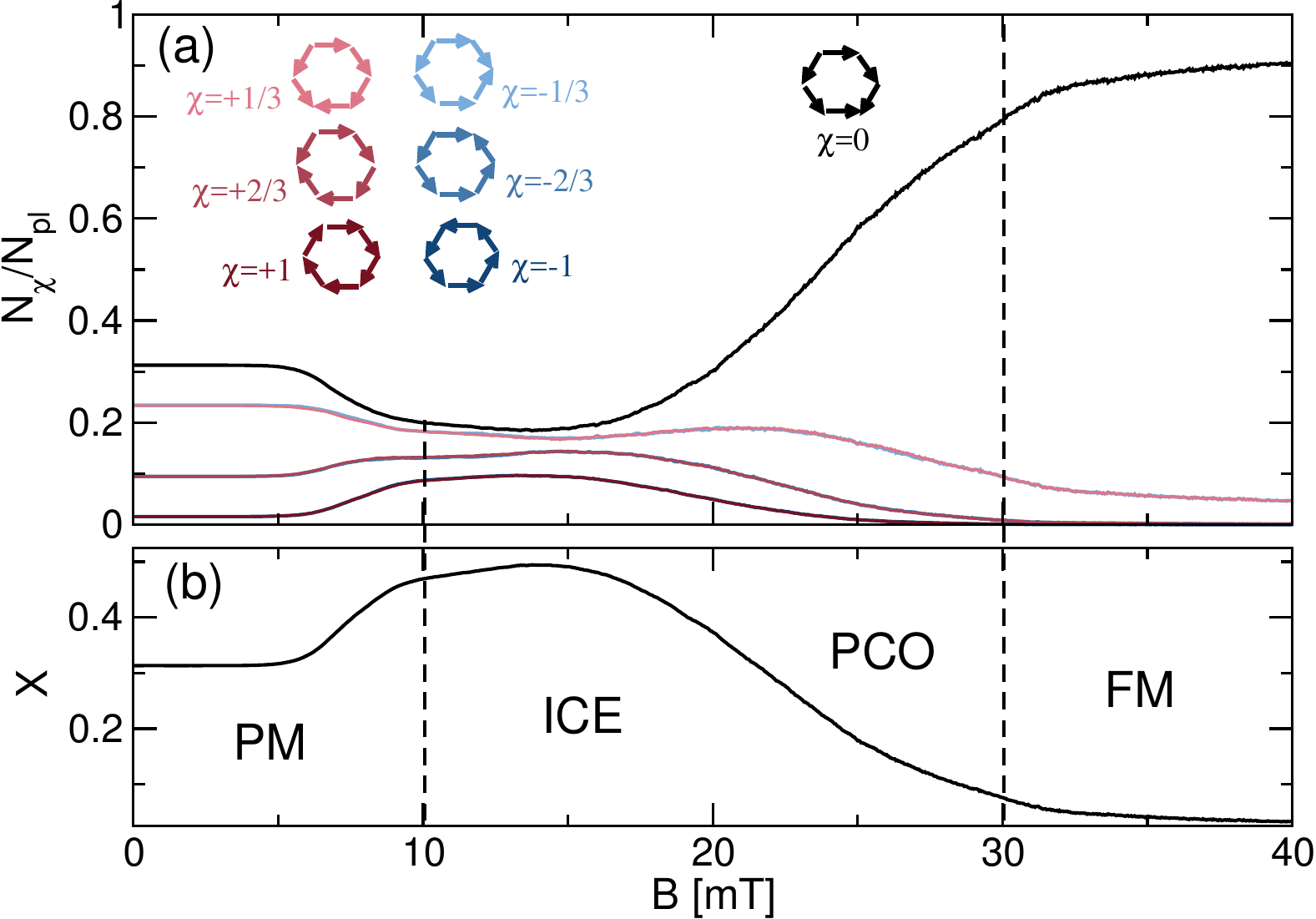}
\caption{(a) Fraction $N_{\chi}/N_{pl}$ of plaquettes with chirality $\chi$
  vs $B$.  Light red: $\chi=\pm 1/3$;
  red: $\chi=\pm 2/3$;
  dark red: $\chi=\pm 1$;
  black: $\chi=0$.
  The following pairs of curves overlap:
  $\chi=\pm 1/3$, $\chi=\pm 2/3$, and $\chi=\pm 1$.
  The chirality first rearranges and then disappears.
  (b) Total chirality $X$ vs $B$. 
}
\label{fig:4}
\end{figure}

We plot the fraction $N_\chi/N_{pl}$ of
hexagonal plaquettes with chirality $\chi$ versus $B$ in Fig.~4(a).
The number of pseudospins with $\chi_i$ aligned in the majority direction is
4 for $\chi=\pm 1/3$,
5 for $\chi=\pm 2/3$,
and 6 for $\chi=\pm 1$, while the
$\chi=0$ plaquettes are achiral.
At the ICE-PCO crossover,
$N_0/N_{pl}$ increases, saturating to
$N_0/N_{pl}\approx 0.9$ in the FM phase.
In Fig.~\ref{fig:4}(b) we plot the total chirality fraction
$X=N_{pl}^{-1}\sum_{i=1}^{N_{pl}}|\chi^{i}|$ versus $B$. 
In the LRO chiral phase in magnetic ASI
\cite{21,22,25}, $X$ increases
from
$X=5/16$ (random) to
$X=2/3$ (LRO). 
In Fig.~\ref{fig:4}(b), $X=5/16$ in the PM phase,
reaches a local maximum in the ICE phase,
and is nearly zero in the FM phase.
These results indicate that the ICE and PCO ice rule obeying phases in
hexagonal colloidal ASI differ in nature
from the SI2
and LRO phases of
magnetic ASI.

\begin{figure}
 \includegraphics[width=\columnwidth]{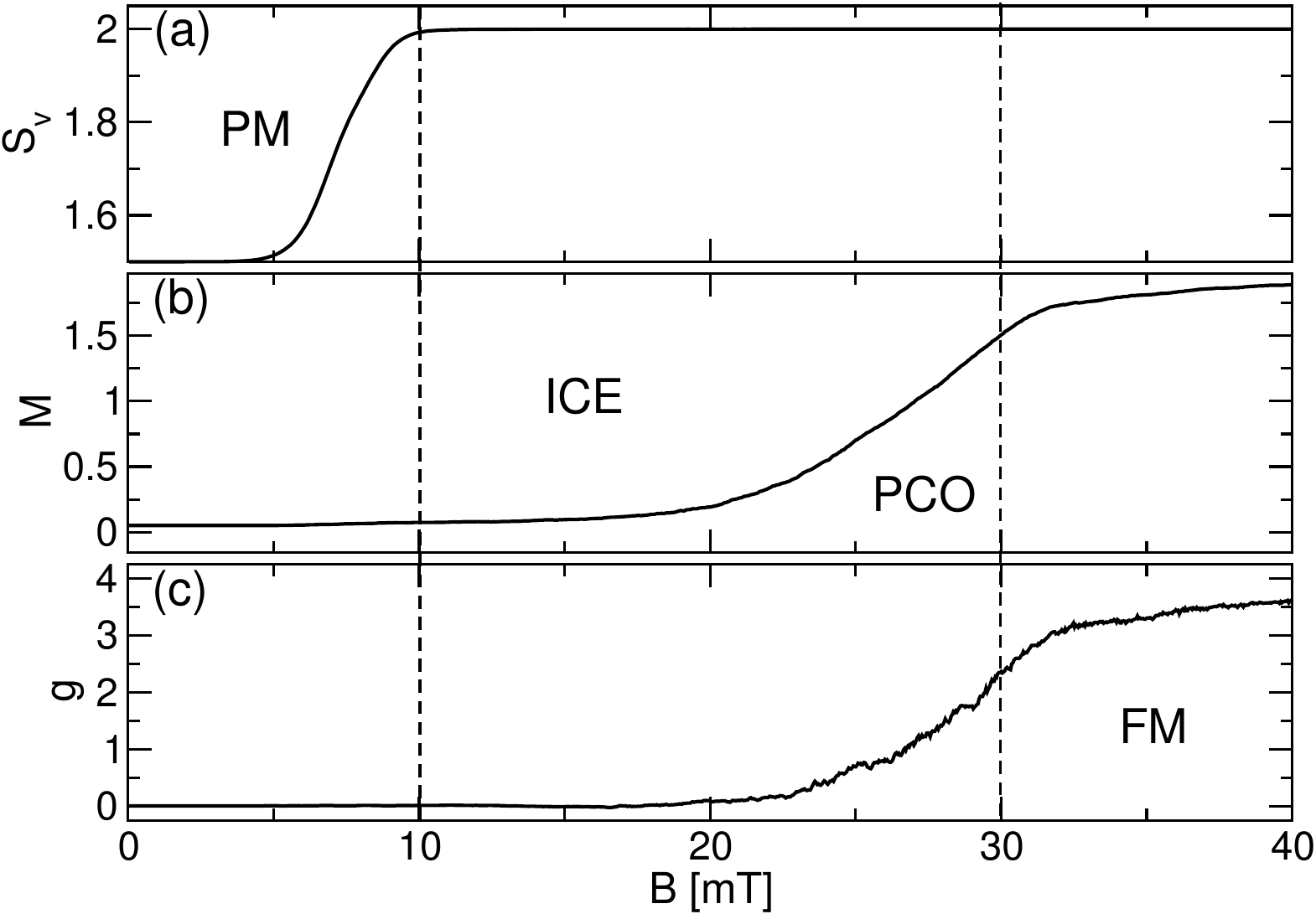}
\caption{(a)
  Average vertex spin $S_v$ vs $B$.
  When $M=2$, all the vertices obey the ice rules.
  (b) Total magnetization $M$ vs $B$
  grows as the system approaches the FM phase.
  (c) Spin-spin correlation $g$ for neighboring vertices vs $B$ 
  increases when FM grains emerge and grow.
}
\label{fig:5}
\end{figure}

To characterize the
FM spin ordering,
we measure the vertex spin-spin correlation for
neighboring vertices,
$g =\langle \vec{s}_i \cdot \vec{s}_j \rangle$, where 
the vertex spin $\vec{s}_i\equiv \sum_{j=1}^3 \vec \sigma_j$,  the sum of the
surrounding pseudospins.
Here, $\vec{s}_i$ points in one of the three
lattice directions and
$|\vec{s}_i|=2$ or 0.
As FM order appears, $g$ increases.
The average vertex spin $S_v = N_v^{-1}\sum_{i=0}^{N_v}|\vec{s_i}|$ 
saturates to $S_v=2$ once all the $q_n=\pm 3$
charges disappear at the PM-ICE transition, as shown
in Fig.~\ref{fig:5}(a).
We plot the total magnetization $M = N_v^{-1}|\sum_{i=0}^{N_v}\vec{s_i}|$ versus $B$
in Fig.~\ref{fig:5}(b).  In the PM and ICE phases, $M=0$,
but above $B=20$ mT, 
$M$ increases, saturating in the FM phase.
The plot of $g$ versus $B$
in Fig.~\ref{fig:5}(c) shows a similar saturation of $g$ in the FM phase.
There are six possible FM orientations, so domains can form as
shown in Fig.~\ref{fig:2}(d) \cite{suppl}, with an average size that
increases as the rate of change of $B$ decreases.
It is possible for the FM domains to be
arranged such that $S_v$ = 0, similar to what is observed in an FM material
that contains ordered domains but has no net magnetization.

{\it Discussion--}
Although an ice manifold forms due to NN vertex energetics,
its inner phases  are typically driven by next-NN interactions~\cite{23}.
In magnetic kagome ASI, an anisotropic dipolar law governs
the long range interactions between in-plane spins,
and it can be shown through multipole expansion that CO
arises from the mutual Coulomb attraction of oppositely charged vertices.
The full dipolar interaction thus produces the
LRO, and the long range interaction tail generates the inner phases.
In contrast, 
our colloids interact isotropically through an inverse-cube repulsion,
and the phases arise not from local energy minimization,
but from an emergent, collective behavior. 

To understand the CO, consider the interaction of two
adjacent ice rule vertices with $n=n_1$ and $n=n_2$.
Approximating the $n$ colloids at each vertex as a composite object located at the
vertex center gives a vertex-vertex interaction energy of
$E_{vv}=n_1n_2 B^2/a_h^3$, which
is minimized when $n_1=n_2=1$, or when each vertex has $q_n=-1$.
The local inter-vertex interaction disfavors the formation of a
CO state, so
both the CO and the ice manifold in colloidal ASI
result from topologically constrained, {\it global}
energy minimization,
since
it is impossible for all the vertices to have $q_n=-1$.
Starting from a CO state
of the type shown in Fig.~\ref{fig:1}(d),
any rearrangement that creates a
pair of $q_n=-1$ charges  lowers the energy locally
by $-B^2/a_h^3$, but also creates a nearby pair of $q_n=2$
charges with a local energy increase of
$3B^2/a_h^3$,
giving a net energy increase of $2B^2/a_h^3$.

The LRO can be viewed as an ordering of emergent dimer spins.
To establish an analogy between our system and the magnetic
ASI, we replace each trap with a double occupancy trap plus a
dumbbell of negative and positive charges,
written symbolically as
\textemdash\tikzcircle{3pt}~$=\frac{1}{2}$\tikzcircle{3pt}\textemdash\tikzcircle{3pt}~$+\frac{1}{2}$\tikzcirclew{3pt}\textemdash\tikzcircle{3pt},
where \tikzcirclew{3pt} represents a ``negative'' colloid.
In the thermodynamic limit, the energetics
are determined by the interaction between the spins
$\vec {\sigma}= $~\tikzcirclew{3pt}\textemdash\tikzcircle{3pt},
exactly as in magnetic ASI, with no contribution from
the double occupancy background.
The LRO differs from that of magnetic ASI
because the dimer spin interaction originates
from the colloidal interactions.
For ideal dipoles this leads to:
$E_{\vec{\sigma}_1,\vec{\sigma}_2}\propto \left[ \vec{\sigma}_1 \cdot\vec{\sigma}_2-5 (\vec{\sigma}_1\cdot \vec{r}_{12})(\vec{\sigma}_2 \cdot \vec{r}_{12})\right]/r_{12}^5$.
This is similar to
the magnetic dipolar interaction, but with a 5
in the exponent and inner coefficient, which
enhances the ferromagnetic coupling and permits the development
of ferromagnetic LRO,
as described by the ``minority spin'' argument of
Ref.~\cite{22} using a different interaction.
If $a_t=a_h$
the colloids  coalesce in the vertex,
producing CO but not LRO.
Thus, the separation of the CO and LRO phases increases
as $a_t/a_h \rightarrow 1$.

{\it Conclusion--}  We observe inner phases within the ice
manifold of hexagonal colloidal
artificial spin ice using a simulation that faithfully mimics experiment.
These phases originate from
interactions between non-nearest-neighbors,
rather than simple vertex energetics, and
disappear for short-ranged interactions,
explaining why they were not observed previously.
Both the inner phases in disordered colloidal systems and
the ice manifold of colloidal artificial spin ice
emerge from global collective behaviors,
rather than from the local energy minimization
found in magnetic kagome ice, producing many additional types of
frustration in the colloidal ice.

\begin{acknowledgments}
This work was carried out under the auspices of the 
NNSA of the U.S. DoE at LANL under Contract No.
DE-AC52-06NA25396.
\end{acknowledgments}

\end{document}